\documentclass[epj,referee]{svjour}
\usepackage{graphics}
\usepackage{psfig}
\newcommand{\dlt}{\bigtriangleup}
\newcommand{\beq}{\begin{equation}}
\newcommand{\eeq}[1]{\label{#1} \end{equation}}
\newcommand{\insertplotlong}[1]{\centerline{\psfig{figure={#1},height=21.5cm}}}
\newcommand{\beqar}{\begin{eqnarray}}
\newcommand{\eeqar}[1]{\label{#1} \end{eqnarray}}

\begin{document}
\title{Non-equilibrated post freeze out distributions}
\author{V.K. Magas\inst{1}\thanks{\emph{Email:} vladimir@cfif.ist.utl.pt }
 \and A. Anderlik\inst{2}\thanks{\emph{Email:} anderlik@fi.uib.no }
\and Cs. Anderlik\inst{3}\thanks{\emph{Email:} csaba@ii.uib.no}
 \and L.P. Csernai\inst{2,4}\thanks{\emph{Email:} csernai@fi.uib.no }
}                     % Do not remove
\institute{Center for Physics of Fundamental Interactions (CFIF)\\
Instituto Superior Tecnico, Av. Rovisco Pais, 1049-001 Lisbon, Portugal
 \and
 Section for Theoretical and Computational Physics, Department of Physics\\
University of Bergen, Allegaten 55, N-5007, Norway
 \and
Parallab - High Performance Computing laboratory\\
University of Bergen, Department of Informatics, Thormehlensgt 55, N-5020 Bergen, Norway
 \and
KFKI Research Institute for Particle and Nuclear Physics\\
P.O.Box 49, 1525 Budapest, Hungary
}
\date{Received: date / Revised version: date}
\abstract{
We discuss freeze out on the hypersurface with time-like normal vector, trying to
answer how realistic is to assume thermal post freeze out distributions
for measured hadrons.
Using simple kinetic models for gradual freeze out we
are able to generate thermal post FO distribution, but only in highly
simplified situation. In a more advanced model, taking into account rescattering and
re-thermalization, the post FO distribution gets more complicated.
The resulting
particle distributions are in qualitative agreement with the experimentally measured
pion spectra.
Our study also shows that the obtained post FO distribution functions, although
analytically very different from the J\"uttner distribution, do look pretty much like thermal
distributions in some range of parameters.
 % end of PACS codes
} %end of abstract
\maketitle
\section {Introduction} \label{int}

Since the very beginning of heavy ion collision physics fluid dynamical models
were widely used for simulation of such reactions.
Their advantage is that one can vary flexibly the Equation of State
(EoS) of the matter and test its consequences on the reaction dynamics
and the outcome. This makes fluid dynamical models a very powerful tool
to study possible phase transitions in heavy ion collisions - such as
the liquid-gas or the Quark-Gluon Plasma (QGP) phase transition.
For highest energies,
achieved nowadays at RHIC, hydrodynamic calculations give a good
description of the observed radial and elliptic flows
\cite{Sollfrank-BigHydro,Schlei-BigHydro,Kolb-UU,Kolb-LowDensity,Kolb-Radial,Htoh}, in
contrast to microscopic models, like HIJING \cite{Molnar-Elliptic} and UrQMD
\cite{UrQMD-Elliptic}.

Freeze out (FO) process is an important and necessary part of the fluid dynamical modeling.
Generally saying, fluid dynamical models are determined by the initial state,
EoS and the way FO is realized, in addition to well known fluid dynamical equations.
FO describes the last stages of the reaction, when the matter becomes so dilute and cold
that particles do not interact  anymore, their momentum distribution freezes out, and
particles stream freely toward the detectors.
FO models allow the evaluation of two-particle correlation data,
different flow analyzes,
transverse momentum and transverse mass spectra and other observables.
There are many different theoretical approaches for the freeze out problem.
Sometimes these approaches are oversimplified to the extent that
conservation of energy and momentum is violated
(see Ref. \cite{FO2,FO3} for an overview).

A basic standard assumption in fluid dynamical modeling of heavy ion collisions
is that freeze out happens across a hypersurface in space-time.
Such a hypersurface is an idealization of a layer of finite
thickness (of the order of at least mean free path or collision time) where the
frozen-out particles are formed and the interactions in the matter become
negligible. The dynamics of this layer is described in different kinetic
models such as Monte Carlo models \cite{BB95,BB97} or four-volume emission
models \cite{BC82,GH95,GH96,GH97,He97}.
Under such an assumption FO can be pictured as a discontinuity in relativistic flow,
where the kinetic
properties of the matter, such as energy density or flow velocity
change suddenly.

It took a long time to develop general theory of discontinuities in the
relativistic flow. The story goes like this.
In 1948 Taub \cite{Ta48} discussed discontinuities across propagating hypersurfaces with
space-like normal vectors. If one applies Taub's formalism
to FO hypersurface with time-like normal vector, one gets a usual
Taub adiabat, but the equation of the Rayleigh line yields imaginary values
for the particle current across the front. Thus, these hypersurfaces were
thought unphysical. However more recently, Taub's approach has been
generalized to these hypersurfaces as well \cite{Cs87} and
the imaginary particle currents arising from the equation of the
Rayleigh line were eliminated.
Thus, it is possible to take into account conservation laws
exactly across any surface of discontinuity in relativistic flow. The corresponding
equations are
\beq
 T_{pre}^{\mu\nu}d\sigma_{\nu} =
 T_{post}^{\mu\nu}d\sigma_{\nu}  \,\, ,~~
\eeq{e1}
\beq
 n_{pre}u_{pre}^{\mu}d\sigma_{\mu}=
 n_{post} u_{post}^{\mu}d\sigma_{\mu} \,\, ,
\eeq{e2}
where $d\sigma ^{\mu}$ is the unit 4-vector normal to the
discontinuity hypersurface,
\beq T^{\mu\nu} = (\epsilon + p)u^{\mu}u^{\nu} -
pg^{\mu\nu}
\eeq{tmn} is the energy-momentum tensor,
\beq
N^\mu=n u^{\mu}
\eeq{nm} is a baryon current
(usually in heavy ion collision  modeling we neglect electric charge current)
These
consist of local thermodynamical fluid quantities: the energy density
$\epsilon$, pressure $p$, baryonic density $n$ and the collective
four-velocity $u^{\mu} = \frac{1}{\sqrt{1-\vec{v}^2}}(1,\vec{v})$.

Since 1974 people were using the Cooper-Frye formula \cite{CF74} to calculate
final particle spectra. Much later it was realized that the Cooper-Frye FO description
has a conceptual problem of negative contributions, when it is applied to the FO
hypersurface with space-like normal vector
\cite{Bu96,FO2,FO1,FO3,FO4,FO5,ref18,bug}.\footnote{
To our knowledge the first attempt to modify post FO distribution
has been done by Yu. Sinyukov in 1989 \cite{Si89}. For the critical
review of this work see \cite{ref18}.}  This leads to the non-equilibrated cut-off
post FO distributions:
\beq
f_{FO}(\vec{r},p;T,n,u^\nu,d\sigma^\mu)
=
f_{FO}(\vec{r},p;T,n,u^\nu) \Theta(p^\mu\ d\sigma_\mu)  \ .
\eeq{cut}
The simplest example of such cut-off distribution is
the cut J\"uttner distribution \cite{Jutt}, proposed in
Ref. \cite{Bu96}. In Refs. \cite{FO1,FO3,FO4,FO5} it was shown that one can obtain
cut J\"uttner post FO distribution in a oversimplified kinetic model. If one improves this model,
taking into account rescattering and re-thermalization, or if one uses more realistic four-volume
emission model \cite{FO3}, then final distributions have more complicated form, although they
are cut-off as required. Unfortunately, so far we do not have good analytical expression
to fit post FO distributions for the hypersurface with space-like (or, generally, any type of)
normal vector. Therefore in most calculations scientists prefer to model FO based on
time-like hypersurfaces. For the ultra-relativistic heavy ion collisions this is justified by
the Bjorken model, where the natural choice of FO hypersurface is $\tau=\tau_{FO}$.

In this paper we ask the following question. Well, we know now that FO through the hypersurface
with space-like normal vector leads to non-equilibrated post FO distributions, but can we really
assume thermal post FO distributions if the FO hypersurface is time-like? This is not restricted by
general theory, but how realistic such an assumption is?
From the experimental data we know that pion and charged hadron (which
are actually pion dominated) transverse mass spectra
both at SPS
and at RHIC \cite{spectra,summspectra} are not the J\"uttner type -
the slope parameter is decreasing with increasing $p_t$.

We generalize the simple model presented in Refs. \cite{FO1,FO3,FO4,FO5}
for the FO hypersurface with time-like normal vector (the model is also not 1D anymore).
The philosophy of this work is also similar - first we will show that in the oversimplified model
the obtained post FO spectra are really thermal ones.
Then in more realistic model, taking into account
rescattering in the still interacting gas, we will obtain post FO distribution
in the more complicated non-equilibrated form, which is in qualitative agreement with experimental data,
mentioned above. Actually
the main goal of this work is not quantitative calculations, but a qualitative
understanding of the phenomenon in simple models.

\section{Simple freeze out model}
\label{fo1}

As it was mentioned above our basic model is in some sense similar to the one
discussed in Refs. \cite{FO1,FO3,FO4,FO5}, but on the other hand it is much more realistic
and also habitual for the people working in
ultra-relativistic heavy ion collision field. Instead of very special geometry (1D, infinitely long tube
with particle source on the left end and vacuum on the right one) used in Refs. \cite{FO1,FO3,FO4,FO5},
this model can be combined with any relativistic hydrodynamical model.
Below we will use proper time coordinate keeping in mind the Bjorken hydrodynamical model, as the
simplest but still reliable one, although the similar treatment could be done for any
pre FO state.

Let assume that at some moment $\tau=\tau_0$ our matter is still completely equilibrated and
described by the thermal distribution $f_{Juttner}(\tau_0,\vec{r},p,T,n,u^\mu)$. Starting from this moment matter
starts to gradually freeze out.
At each particular moment $\tau_i$ the FO hypersurface will be $\tau=\tau_i$, and we will follow
this hypersurface, i.e. we will describe our system from the reference frame of the front
(RFF), where normal vector is $d\sigma_\mu=(1,0,0,0)$.
We can describe the FO kinetics of such a system assuming that we
have two components of our distribution \cite{GH95,GH96,GH97,FO1,FO3,FO4,FO5}:
$f_{free}(\tau,\vec{r},p)$, describing frozen out particles, and
$f_{int}(\tau,\vec{r},p)$, describing particles, which still interact
(below we will use shorter notations, $f_{free}(\tau)$ and $f_{int}(\tau)$,
but the meaning is the same).
At the initial moment, $\tau=\tau_0$, the distribution  $f_{free}$ vanishes exactly and
$f_{int}(\tau_0)=f_{Juttner}(\tau_0,\vec{r},p,T,n,u^\mu)$.
In this work we will use an extended definition of the classical J\"uttner distribution \cite{Jutt},
which is adequate at ultra-relativistic energies and allow for particle creation. Thus
the particle density, which depends on $T$, is not the same as the conserved baryon charge density, $n$.
Then, $f_{int}$
gradually disappears and $f_{free}$ gradually builds up as $\tau$ increases.
The most simple kinetic model describing the
evolution of such a system is the following:
$$
\partial_\tau f_{int}(\tau)   = - \frac{1}{\tau_f}
           f_{int}(\tau)\, ,
$$
\beq
\eeq{kin-1}
$$\partial_\tau f_{free}(\tau)   = \frac{1}{\tau_f}
           f_{int}(\tau)\, ,
$$
where $\tau_f$ is a parameter with dimension of time determining how fast
particles freeze out. It is of the order of collision time. In general case
$\tau_f$ may depend on particle density and momentum of the particle, but in our
simple model we neglect such effects and consider $\tau_f=const$. Our assumption becomes particularly bad for the large times, when the particle density is very low - we will discuss this problem separately in section
\ref{pi}.  If this approximation is not done, i.e. the relaxation time is assumed to be momentum and density dependent, that would be a large step towards the full solution of the BTE, which is then a much more involved problem, and in most cases cannot be performed analytically. Nevertheless even this strongly simplifying assumption leads to  nontrivial results as we shall see later.

The interacting component, $f_{int}$, will gradually freeze out
in an exponential rarefaction with time.
\beq
f_{int}(\tau) =  f_{Juttner}(\tau_0) e^{-\frac{\tau-\tau_0}{\tau_f}}.
\eeq{sol-11}
Then
\beq
f_{free}(\tau) =   f_{Juttner}(\tau_0) \left(1-
e^{-\frac{\tau-\tau_0}{\tau_f}}\right)\,.
\eeq{sol-12}
At $\tau \longrightarrow \infty$ the distribution $f_{free}$  will tend to the original
J\"uttner distribution, while $f_{int}$ will vanish. The frozen out particles keep their velocities
and consequently the volume occupied by the frozen out matter will gradually increase with time.

This is a highly unrealistic model, similar to the one resulting
in the cut J\"uttner distribution in Refs. \cite{FO1,FO3,FO4,FO5} for the FO hypersurface
with space-like normal vector.  We can improve this model taking into account
rescattering in the interacting component in the same way as
it was done in the References above.

\subsection{Freeze out model with rescattering}
\label{fo-re}

The assumption that the interacting part of the distribution, $f_{int}$,
remains after some drain the
same J\"uttner distribution with decreased amplitude is, of course, highly
unrealistic.  As suggested in Refs. \cite{FO1,FO3,FO4,FO5} rescattering within this
component will lead to re-thermalization and \linebreak re-equilibration of this
component. Thus, the evolution of the component $f_{int}$ is determined both by
the drain term and the re-equilibration.
If we include the collision terms explicitly into the transport equations
(\ref{kin-1}) this in general case leads to a combined set of integro-differential
equations. We can, however, take advantage of the  relaxation
time approximation to simplify the description of the dynamics.
In this framework
the two components of the momentum distribution develop according to the
modified coupled differential equations:\footnote{Notice that in the ultra-relativistic
case the total number of particles is much bigger than the baryon charge and depends on
$T$. Thus the collision term in system (\ref{kin-2}) gives a
nontrivial and nonvanishing contribution.}
%\pagebreak
$$
\partial_\tau f_{int}(\tau)   = - \frac{1}{\tau_f}
           f_{int}(\tau)+\left[ f_{eq}(\tau) -  f_{int}(\tau)\right]
           \frac{1}{\tau_r}\,,
$$
\beq
\eeq{kin-2}
$$
\partial_\tau f_{free}(\tau)  = \frac{1}{\tau_f}
           f_{int}(\tau)\, .
$$

Thus, the interacting component of the momentum distribution, described by
eq. (\ref{kin-2}), shows the tendency to approach an equilibrated
distribution with a relaxation time  $\tau_r $.  Of course
due to the energy, momentum and conserved particle drain, this distribution,
$f_{eq}(\tau)$ is not the same as the initial J\"uttner distribution,
but its parameters, $n_{eq}(\tau)$, $T_{eq}(\tau)$ and $u^\mu_{eq}(\tau)$,
change as required by the conservation laws.

If we do not have collision or relaxation terms in our transport equation
then the conservation laws are trivially satisfied, because the change of the full
distribution function is
$\partial_\tau f=\partial_\tau f_{int}(\tau)  + \partial_\tau f_{free}(\tau)=0$.
If, however, collision or
relaxation terms are present, these contribute to the change of $T^{\mu\nu}$
and $N^\mu$, and this should be considered in the modified
distribution function $f_{int}(\tau)$.
In this case from the conservation laws we get:
\beq
d N_{int}^\mu = - d N_{free}^\mu =
      -\frac{d\tau}{\tau_f}\int\frac{d^3p}{p_0} p^\mu
   f_{int}(\tau)=-\frac{d\tau}{\tau_f}N_{int}^\mu
\eeq{kin-4i}
and
\beq
d T_{int}^{\mu\nu} = - d T_{free}^{\mu\nu} =
      -\frac{d\tau}{\tau_f}\int\frac{d^3p}{p_0} p^\mu p^\nu f_{int}(\tau)
      =-\frac{d\tau}{\tau_f}T_{int}^{\mu\nu}\,,
\eeq{kin-5i}
where we used kinetic definitions of $N^\mu$ and $T^{\mu\nu}$.
The above equations can be easily solved:
\beq
N_{int}^\mu(\tau) = N^\mu(\tau_0)
      e^{-\frac{\tau-\tau_0}{\tau_f}}
\eeq{kin-4}
and
\beq
T_{int}^{\mu\nu}(\tau) = T^{\mu\nu}(\tau_0)
      e^{-\frac{\tau-\tau_0}{\tau_f}}\,.
\eeq{kin-5}

\subsection{Immediate re-thermalization limit}
\label{ss1}

As a first approximation to this solution let us assume
that $\tau_r \ll \tau_f$, i.e. re-thermalization is much faster
than particles freezing out, or much faster than parameters,
$n_{eq}(\tau)$, $T_{eq}(\tau)$ and $u^\mu_{eq}(\tau)$ change.
This leads to
\beq
f_{int}(\tau)\approx f_{eq}(\tau), \ \ {\rm for} \ \
\tau_r \ll \tau_f \,.
\eeq{kin-7}
This assumption may be unrealistic for large times, when particle density becomes very
low. We shall discuss the problem of the large times in the next section.

For $f_{eq}(\tau)$ we can assume the usual J\"uttner form at
any $\tau$ with parameters $n(\tau),\ T(\tau)$ and $u^\mu(\tau)$, where
$u^\mu(\tau)$
is the flow velocity in the RFF frame.
Under such an assumption the interacting component is all the time in equilibrium, and
therefore baryon current and energy momentum tensor of this component have the usual
forms, eqs. (\ref{tmn},\ref{nm}). Thus, we can write eqs. (\ref{kin-4},\ref{kin-5}) in
the following forms:
\beq
n(\tau)u^\mu(\tau) = n(\tau_0)u^\mu(\tau_0)
      e^{-\frac{\tau-\tau_0}{\tau_f}}
\eeq{kin-4a}
and
$$
(\epsilon(\tau)  + p(\tau) )u^{\mu}(\tau) u^{\nu}(\tau)  -
p(\tau) g^{\mu\nu}=
$$
\beq
=\left[(\epsilon(\tau_0) + p(\tau_0))u^{\mu}(\tau_0)u^{\nu}(\tau_0) -
p(\tau_0)g^{\mu\nu}\right]e^{-\frac{\tau-\tau_0}{\tau_f}}\,.
\eeq{kin-5a}

Using the expressions derived in Ref. \cite{FO3} we can easily find the
solutions for $n(\tau),\ \epsilon(\tau),\ p(\tau)$ and $u^\mu(\tau)$. First of all, it is nice
to see that now we do not have a problem with the flow velocity definition (for the FO model discussed in
Refs.\cite{FO1,FO3,FO4,FO5} the Landau and Eckart definitions of the flow velocity were
giving different results), because the velocity change vanishes:
$$
du^\mu_{Eckart}(\tau)=\Delta^{\mu\nu}(\tau)\  \frac{dN_{\nu}(\tau)}{n(\tau)}=
$$
\beq
=du^\mu_{Landau}(\tau) = \frac{
\Delta^{\mu\nu}(\tau)\ \ dT_{\nu\sigma}(\tau)\ \ u^\sigma(\tau)}{\epsilon(\tau) + P(\tau)}=0\,,
\eeq{u}
where
$
\Delta^{\mu\nu}(\tau) = g^{\mu\nu} -  u^\mu(\tau)\, u^\nu(\tau)\,
$\ \
is a projector to the plane orthogonal to
$
u^\mu(\tau)
$. Thus,
\beq u^\mu(\tau)=u^\mu(\tau_0)=u^\mu=const\,.
\eeq{u2}

Then,
\beq
d\epsilon(\tau) =u_{\mu}\ dT^{\mu\nu}_{int}(\tau)\ u_{\nu} \, \rightarrow
\epsilon(\tau)=\epsilon(\tau_0)e^{-\frac{\tau-\tau_0}{\tau_f}}\,.
\eeq{etau}
From eqs. (\ref{kin-4a},\ref{u2}) we obtain the expression for the baryon density
\beq
n(\tau)=n(\tau_0)e^{-\frac{\tau-\tau_0}{\tau_f}}\,.
\eeq{ntau}
Similarly, from eqs. (\ref{kin-5a},\ref{u2},\ref{etau}) we obtain
\beq
p(\tau)=p(\tau_0)e^{-\frac{\tau-\tau_0}{\tau_f}}\,.
\eeq{ptau}
This last expression means that from the moment our gradual FO has started the pressure in the interacting
component is fixed by the model and conservation laws, and one does not need in addition the EoS as usual.
The EoS during FO looks like:
\beq
p(\tau)=\frac{p(\tau_0)}{\epsilon(\tau_0)}\epsilon(\tau)\,.
\eeq{pEoS}

\subsection{Freeze out of massless pion gas}
\label{pi}

To proceed further we need to make some assumptions about the type of matter we are dealing with.
First of all we assume that our system is homogeneous (this assumption will be
used for the next sections as well).
Let us take the simplest case of massless pion gas. Then, EoS is
\beq
\epsilon=\sigma_{SB}T^4 \,,
\eeq{SB}
\beq
p=\frac{1}{3}\epsilon\,,
\eeq{plasma}
and it will stay like this during all the FO process.
Using eq. (\ref{etau}) we find temperature as a function of $\tau$:
\beq
T(\tau)=T(\tau_0)e^{-\frac{\tau-\tau_0}{4\tau_f}}\,.
\eeq{Ttau}
Thus, we can write the expression for the Maxwell-Boltzmann (MB)
distribution function of the interacting component (we are working in the system, where
$\hbar=c=1$ and assume for simplicity only one type of pions. i.e. degeneracy $g=1$):
\beq
f_{int}(\tau)=\frac{1}{(2\pi)^3}e^{-\frac{p^\mu u_\mu}{T(\tau)}}=
\frac{1}{(2\pi)^3}e^{-\frac{p^\mu u_\mu}{T(\tau_0)e^{-\frac{\tau-\tau_0}{4\tau_f}}}}\, .
\eeq{inttau1}
\beq
f_{int}(\tau)|_{\tau \rightarrow \infty} \rightarrow 0\, ,
\eeq{intinf1}
as it should. For such a distribution
\beq
\sigma_{SB}=\frac{3}{\pi^2} \,.
\eeq{SB2}

Now, based on the second equation in  system (\ref{kin-2}),
 we can find the distribution
function of the frozen out particles.
\beq
f_{free}(\tau)=\frac{4}{(2\pi)^3}
\left[Ei\left(-\frac{p^\mu u_\mu}{T(\tau_0)}e^{\frac{\tau-\tau_0}{4\tau_f}}\right)
- Ei\left(-\frac{p^\mu u_\mu}{T(\tau_0)}\right)\right]\,,
\eeq{freetau1}
where
\beq
Ei(y)=\int_{-\infty}^{y}\frac{e^{y}}{y}dy\, ,
\eeq{Ei}
is the Exponential integral function.
\beq
f_{free}(\tau)|_{\tau \rightarrow \infty} \rightarrow
-\frac{4}{(2\pi)^3}Ei\left(-\frac{p^\mu u_\mu}{T(\tau_0)}\right)\, ,
\eeq{freeinf1}

\begin{figure*}[htb]
%\vspace*{-3.5cm}
        \insertplotlong{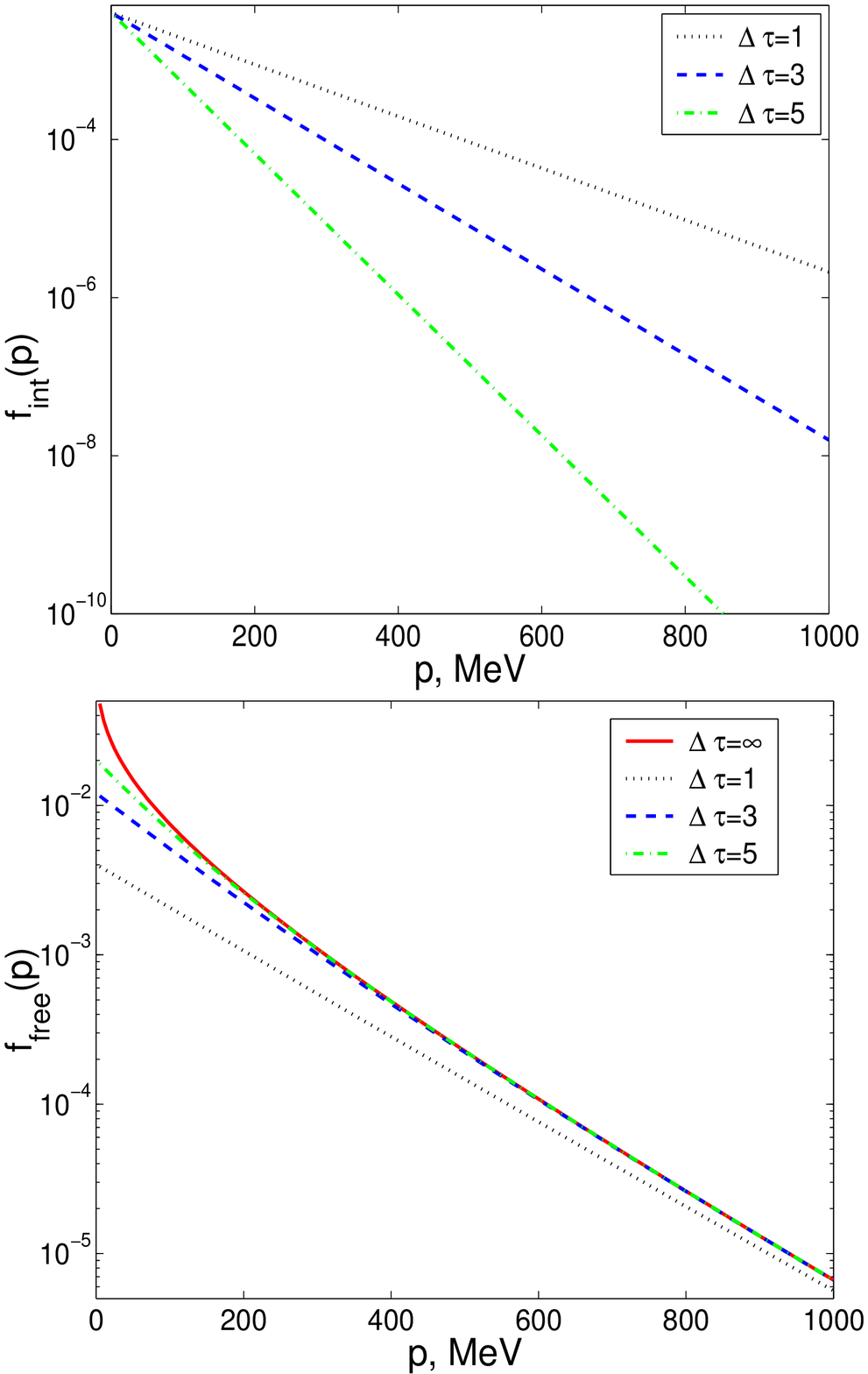}
\vspace*{-0.5cm}
\caption{Figure shows the evolution of the $f_{int}$ (upper subplot)
and $f_{free}$ (lower subplot) with time, $\dlt \tau=(\tau-\tau_0)/\tau_f$. The initial
flow velocity (which is actually kept all the time) is $v=0$, $T(\tau_0)=170\ MeV$.
We present distribution functions vs. absolute value of the momentum,
since for zero flow velocity all the directions are equivalent.
}
%\vspace*{-0.6cm}
\label{f1}
\end{figure*}

Figure \ref{f1} shows the evolution of the $f_{int}$ (upper subplot)
and $f_{free}$ (lower subplot) with time, $\dlt \tau=(\tau-\tau_0)/\tau_f$. The initial
flow velocity (which is actually kept constant all the time) is $v=0$ (inspired by Bjorken model).
$T(\tau_0)=170\ MeV$ - we want to start our simulation right after hadronization.
We present the distribution
functions vs. absolute value of the momentum, as for zero flow velocity all the
directions are equivalent.

We would like to underline the following points about these graphs. One may notice that
for the small times $\tau$ ($\dlt \tau\le 3$) $f_{free}$ looks pretty similar to some
thermal distribution - a straight line in the logarithmic scale in our case - although the analytical
expression is very different. For larger $\dlt \tau$ values $f_{free}$ starts to show clear deviations
from the thermal-like distribution. Its non-thermal behaviour, namely the decrease of the
slope with increasing momentum, is in qualitative agreement with the experimentally measured
pion and charged hadrons spectra \cite{spectra,summspectra}.

For the very large times our distribution becomes
very strongly peaked at low momenta. This is not a surprise, since our
temperature is going down exponentially and for large times we keep only particles
with very low momenta in our interacting component, which keeps gradually freezing out
and cooling down.

As it was mentioned before our model becomes unrealistic for large times, but
there is staightforward way to solve this problem.
After some time the particle density will be so low that the particles
will have no chance for further interaction.
So, we have to stop at some moment and transfer
the rest of still "interacting" particles (probably a few percents)
to frozen out component. In any case, since our FO is a very fast exponential process,
the uncertainty generated at large times will be rather small.

\subsection{Entropy condition}
\label{entrp}
To obtain a physically realizable result, we have to check the
condition for entropy increase:
\begin{equation}
S_{\tau} - S_{\tau_0} \ge 0\,,  \label{efo9}
\end{equation}
where, for both the equilibrium and the non-equilibrium distributions
the entropy current is given as
\beq
S^\mu =s u^\mu=-\int \frac{d^3p}{p^0}\ p^\mu \ f(\tau)\ \left[ \ln \left\{
\left( 2\pi \right) ^3\ f(\tau)\right\} -1\right]  \,.
\eeq{scalc}
This condition is not necessary to obtain a solution of the freeze out
problem, but it should always be checked to exclude non-physical solutions.

In our case the original entropy density is given by the well known expression for the
Maxwell-Boltzmann gas:
\beq
s(\tau_0)=\frac{4}{3}\frac{\epsilon(\tau_0)}{T(\tau_0)}=\frac{4}{\pi^2}T^3(\tau_0)\,,
\eeq{s0}
where we have used eq. (\ref{SB2}).
The entropy density for $\tau=\infty$, when all the particles are frozen out, can be calculated
through eq. (\ref{scalc}) using the distribution function given by expression (\ref{freeinf1}).
We have done this integration  numerically and obtained for the ratio of entropy densities:
\beq
R=\frac{s(\tau=\infty)}{s(\tau_0)}=0.9449\,,
\eeq{R}
for any initial temperature, $T(\tau_0)$\footnote{It is not a surprise that $R$ is independent
on $T(\tau_0)$, since for the massless gas we have only one dimensional parameter - $T(\tau_0)$ and
correspondingly $s\propto T^3(\tau_0)$. }.

Thus, we see that the entropy density is slightly decreasing during our FO process. Since we have
a gradual FO this means, that in order to avoid a decrease of total entropy, our FO model
has to be
accompanied by an expansion, which is natural, as mentioned before, and
which should lead to a volume increase by at least $6\%$ during FO.
This does not sound unrealistic.

\subsection{Freeze out of nucleons}
\label{nucl}

Now we would like to study the situation for baryon rich matter - in this
case the distribution functions will be different, since we have to take
care about baryon density conservation in addition.
If our system contains some nucleons in addition to pions (for simplicity we assume
that there are no other mesons in the system), then the situation
is the following.
The EoS at the pre FO state was more complicated then eq. (\ref{plasma}), but for
the FO times it is fixed by the eq. (\ref{pEoS}). We assume that the temperature is
high and the chemical potential is not too large - such that our system is pion dominated
(this is reasonable assumption for ultra-relativistic heavy ion collisions). For
such a system the temperature will be still basically defined by pions and we can
connect $\epsilon$ and $T$ through equation (\ref{SB})
and correspondingly eq. (\ref{Ttau}) is also valid.\footnote{
If we would use Bose and Fermi statistics
 we could formulate this approach in the following way:
\beq
\epsilon=\sigma_{SB}T^4 + O\left(\frac{\mu^2}{T^2}\right)\,,
\eeq{nEoS}
 and
\beq
n=a_{SB}T^2\mu+O\left(\frac{\mu^2}{T^2}\right)\,.
\eeq{nEoSn}
From the last equation using eqs. (\ref{Ttau},\ref{ntau}), we can find that
\beq
\mu(\tau)=\mu(\tau_0)e^{-\frac{\tau-\tau_0}{2\tau_f}}\,,
\eeq{mutau}
and, thus, $\mu$ is decreasing faster than $T$, and therefore if the inequality
$\mu\ll T$, holds for $\tau=\tau_0$, it will be even stronger for higher $\tau$.}

The expression for the
distribution function of the interacting component now becomes (see for example \cite{Csbook}):
\beq
f_{int}(\tau)=\frac{1}{2\pi m^2 T K_2(m/T)}\ n(\tau)e^{-\frac{p^\mu u_\mu}{T(\tau)}}\, ,
\eeq{inttau2a}
where $m$ is the mass of the nucleon, $K_2(x)$ is the Bessel
function of the second kind.
Of course, for the proper treatment of the nucleons we would have to use Fermi statistics,
but in this work we restrict ourselves to MB distributions first of all to
get an analytical/semianalytical results, and secondly because MB
distributions are nevertheless frequently
used to fit experimental data, see for example \cite{summspectra}.
Since for the FO in heavy ion collisions $T\ll m$, we can use the expansion of the Bessel function:
$$K_n(z)|_{z \rightarrow \infty} =
\sqrt{\frac{\pi}{4z}}e^{-z}\left(1+O\left({1\over z}\right)\right)\,.$$
Thus we obtain
$$
f_{int}(\tau)= C T^{-3/2}(\tau)n(\tau)e^{-\frac{p^\mu u_\mu-m}{T(\tau)}}=
$$
\beq
= C\ T^{-3/2}(\tau_0) n(\tau_0)e^{-{5\over 8}\frac{\tau-\tau_0}{\tau_f}}
e^{-\frac{p^\mu u_\mu-m}{T(\tau_0)e^{-\frac{\tau-\tau_0}{4\tau_f}}}}\, ,
\eeq{inttau2}
where $C=\frac{1}{(2m\pi)^{3/2}}$. Again we see that
$
f_{int}(\tau)|_{\tau \rightarrow \infty} \rightarrow 0\, .
$

Now, based on the second equation in system (\ref{kin-2}),
we can calculate the distribution
function of the frozen out particles.
$$
f_{free}(\tau)=4 C T^{-3/2}(\tau_0) n(\tau_0) \left(p^\mu u_\mu-m\right)^{5/2}\times
$$
\beq
\times \left[\Gamma\left(-\frac{5}{2},p^\mu u_\mu-m\right)
-\Gamma\left(-\frac{5}{2},(p^\mu u_\mu-m) e^{\frac{\tau-\tau_0}{4\tau_f}}\right) \right]\,,
\eeq{freetau2}
where
\beq
\Gamma(\alpha,x)=\int_{x}^{\infty}e^{-x}x^{\alpha-1}dx
\eeq{gam}
is the incomplete gamma function.\footnote{Please note that the Exponential integral function, eq. (\ref{Ei}),
is actually a special case of incomplete gamma function:
$\Gamma(0,x)=-Ei(-x)$.}
$$
f_{free}(\tau)|_{\tau \rightarrow \infty} \rightarrow
$$
\beq
4 C T^{-3/2}(\tau_0) n(\tau_0) \left(p^\mu u_\mu-m\right)^{5/2}
\Gamma\left(-\frac{5}{2},p^\mu u_\mu-m\right)\, .
\eeq{freeinf2}

\begin{figure*}[htb]
%\vspace*{-3.5cm}
        \insertplotlong{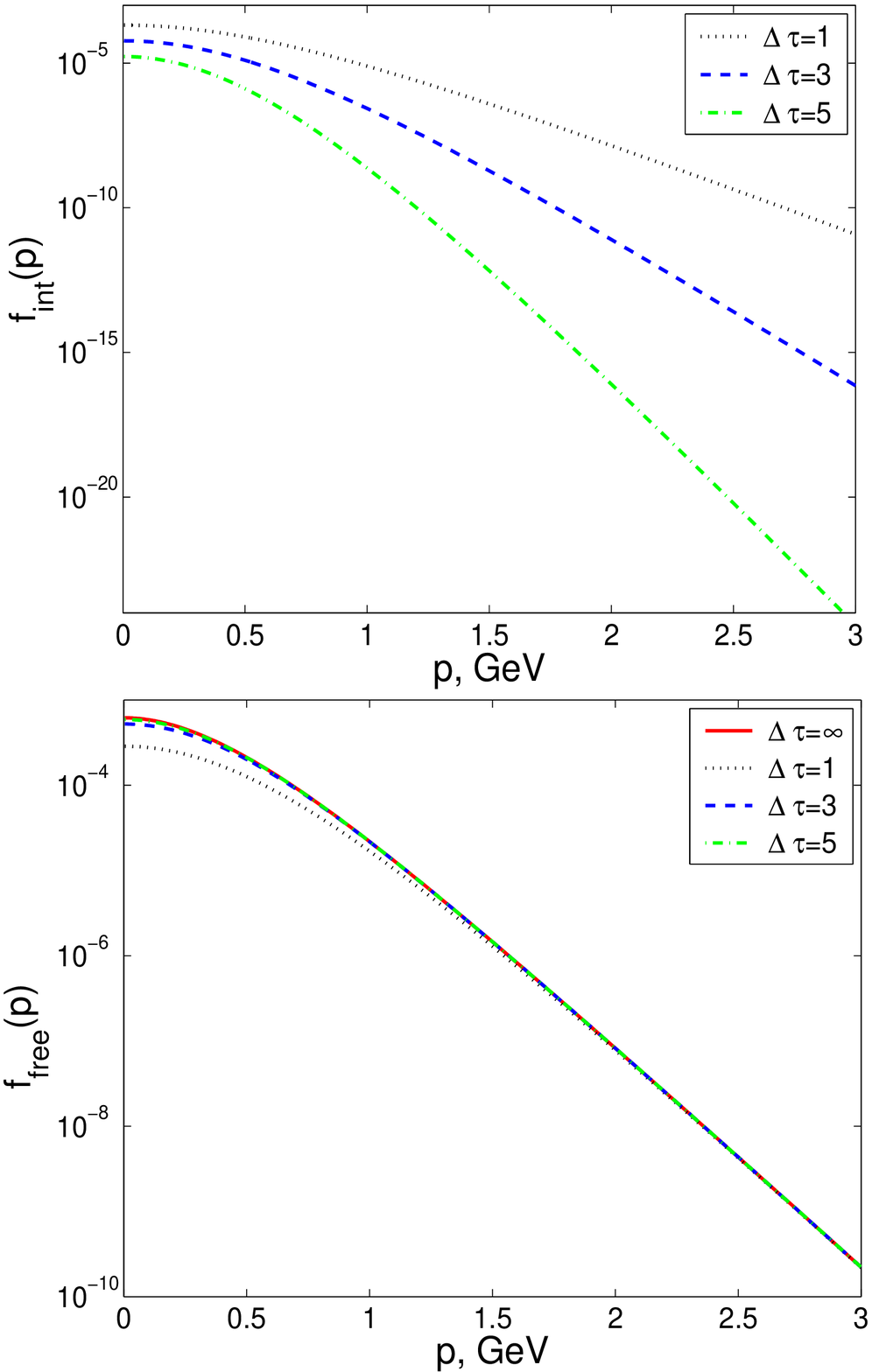}
\vspace*{-0.5cm}
\caption{The same as Fig. \ref{f1}, but for the nucleon distribution functions. Calculations
are done for $n(\tau_0)=0.0025\ fm^{-3}$.
}
%\vspace*{-0.6cm}
\label{f2}
\end{figure*}

The nucleon distribution functions are presented in Figure \ref{f2}.
We choose the initial baryon density much smaller than normal nuclear density,
$n(\tau_0)=0.0025\ fm^{-3}$, to justify our approach \cite{Csbook} --
$$\mu(\tau_0)=T(\tau_0)\ln\left(\left(\frac{2\pi}{mT(\tau_0)}\right)^{3/2}n e^{m/T(\tau_0}
\right)=$$
$$=31\ MeV\ll T(\tau_0)=170\ MeV\,.$$
Again
$f_{free}$ looks similar to some
thermal distribution. In this case $f_{free}$ is saturating very fast - already the
$f_{free}$ $(\dlt \tau =3)$
is a good approximation of the final $f_{free}(\dlt \tau =\infty)$. This is due to the fact
that $f_{int}$ is freezing out much faster for baryons since, in addition to the exponentially
decreasing temperature, it now has the baryon density, $n(\tau)$, in front,
which drops down even faster than $T$.

\section{Freeze Out of dynamically expanding system}
As it was shown in section \ref{entrp} we have to apply our FO model to
the expanding system, what is natural in heavy ion collisions. The simplest, but still nontrivial,
choice we kept in mind from the very beginning is the Bjorken hydrodynamical model.

Combination of this FO scenario with Bjorken scenario is straightforward. Let us, for
example, concentrate on the energy density. The initial state is given at some $\tau_{in}$:
$\epsilon=\epsilon(\tau_{in})$. From this moment energy density can be calculated in the
Bjorken hydrodynamical model from the equation:
\beq
\frac{\partial \epsilon}{\partial \tau}= - \frac{\epsilon+p}{\tau}\,.
\eeq{bjor}
For the ideal ultra-relativistic gas with EoS $p=\epsilon/3$ we obtain:
\beq
\epsilon(\tau)=\epsilon(\tau_{in})\left(\frac{\tau_{in}}{\tau}\right)^{4/3}\,.
\eeq{bjor2}
Eq. (\ref{bjor2}) is valid until moment $\tau=\tau_0$, when FO starts. Then during FO process
energy density is given by the expression:
\beq
\epsilon(\tau)=\epsilon(\tau_0)e^{-\frac{\tau-\tau_0}{\tau_f}}=
\epsilon(\tau_{in})\left(\frac{\tau_{in}}{\tau_0}\right)^{4/3}e^{-\frac{\tau-\tau_0}{\tau_f}}\,.
\eeq{bjor3}

\section{Conclusions}
\label{con}

Our analysis shows that, although the
thermal post FO distributions  are not restricted by
general theory for the FO hypersurface with time-like normal vector,
it may be too optimistic to fit particle spectra with thermal distributions.
In our simple kinetic models we  are able to calculate post FO distributions analytically,
 and we may indeed reproduce thermal post FO distribution, but only in highly
 simplified situation. If we take into account rescattering and consequently re-thermalization
in the interacting component, the post FO distribution gets a more complicated, non-thermal form.
The resulting
particle distributions are in qualitative agreement with the experimentally measured
pions spectra.

To clarify the principal place of our work in the field we shall make a comparison with the cut J\"uttner approach of Bugaev \cite{Bu96}, eq. \ref{cut}, which is the only other post FO distribution published up to recently, which satisfies the FO requirements for space-like FO surfaces. Afterwards, Bugaev's approach was improved for space-like surfaces in Refs. \cite{FO1,FO3} and in subsequent publications by this research group. The present work completes this improvement work and extends it to time-like FO hypersurfaces also. Thus, more precisely, this approach is an improvement compared to Ref. \cite{Bu96} just as the works \cite{FO1,FO3} . This work actually completes the approach introduced in \cite{FO1,FO3} by extending them to  time-like freeze-out surfaces.  The approach is not the best of all possible approaches, it includes a few simplifying assumptions for the more transparent presentation, but it is an important advance compared to the cut J\"uttner approach.

Another important point one can learn from our study is that
although our post FO distribution functions are analytically
very different from the J\"uttner distribution, they
do look pretty much like thermal
distributions in some range of their parameters.
Thus, observed spectra may easily be misinterpreted.

\section*{Acknowledgments}
One of authors, V.M., acknowledge the support of the
Bergen Computational Physics Laboratory in the framework of the
European Community - Access to Research Infrastructure action of the
Improving Human Potential Programme.

%
% BibTeX users please use
% \bibliographystyle{}
% \bibliography{}

\end{document}